\documentclass{article}

\usepackage[english]{babel}
\usepackage{subfig}
\usepackage[letterpaper,top=2cm,bottom=2cm,left=3cm,right=3cm,marginparwidth=1.75cm]{geometry}
\usepackage[numbers,super,compress]{natbib}
\newcommand{\bm}[1]{\mbox{\boldmath $#1$}}

\usepackage{amsmath}
\usepackage{amssymb}
\usepackage{graphicx}
\usepackage[colorlinks=true, allcolors=blue]{hyperref}
\usepackage{algorithmic}
\usepackage{algorithm}
\usepackage{mathtools}
\usepackage{hyperref}

\title{An approach to design adaptive clinical trials with time-to-event outcomes based on a general Bayesian posterior distribution}
\author{James M. McGree$^{1,*}$, Antony M. Overstall$^2$, Mark Jones$^3$ and Robert K. Mahar$^{4,5}$\\
~ \\
$^*$Corresponding author (email: \texttt{james.mcgree@qut.edu.au}) \\
2 George Street, Brisbane, Queensland 4001, Australia \\
~ \\
$^1$School of Mathematical Sciences \\
Queensland University of Technology \\
Brisbane, Australia \\
~ \\
$^2$School of Mathematical Sciences \\ University of Southampton\\ Southampton, United Kingdom \\
~ \\
$^3$Sydney School of Public Health \\
University of Sydney \\
Sydney, Australia \\
~ \\
$^4$Melbourne School of Population and Global Health \\
University of Melbourne \\
Melbourne, Australia \\
~ \\
$^5$Clinical Epidemiology and Biostatistics Unit \\
Murdoch Children's Research Institute \\
Parkville, Australia}

\begin{document}
\maketitle

\begin{abstract}
Clinical trials are an integral component of medical research. 
Trials require careful design to, for example, maintain the safety of participants and to use resources efficiently. Adaptive clinical trials are often more efficient and ethical than standard or non-adaptive trials because they can require fewer participants, target more promising treatments, and stop early with sufficient evidence of effectiveness or harm. 
The design of adaptive trials is usually undertaken via simulation which requires assumptions about the data generating process to be specified {\it a priori}. 
Unfortunately, if such assumptions are misspecified, then the resulting trial design may not perform as expected, leading to, for example, reduced statistical power or an increased Type I error.
Motivated by a clinical trial of a vaccine to protect against gastroenteritis in infants, we propose an approach to design adaptive clinical trials with time-to-event outcomes without needing to explicitly define the data generating process.
To facilitate this, we consider trial design within a general Bayesian framework where inference about the treatment effect is based on the partial likelihood.  
As a result, inference is robust to the form of the baseline hazard function, and we exploit this property to undertake trial design when the data generating process is only implicitly defined. 
The benefits of this approach are demonstrated via an illustrative example and via  redesigning our motivating clinical trial.
\end{abstract}


\section{Introduction}

Efficient clinical trials are needed to accelerate access to new treatments and to reduce the financial and participant burden of medical research. 
Adaptive trials are particularly appealing in this regard because, based on the analysis of accruing data, without compromising the validity of the trial \cite{chow_chang_pong_2005}: (1) the trial can stop early due to success; (2) treatments can be dropped from the trial if they are deemed ineffective or unsafe; (3) new treatments can be included mid-trial as new therapies emerge; and (4) treatments can be randomised such that the most promising are allocated with higher probability.\cite{pallmann_2018} 
Consequently, when compared to fixed or inflexible trial designs, adaptive trials can complete sooner, cost less to run, and reduce the number of participants on inferior treatments.\cite{fda_2010,thorland_2018} 
In designing such trials, rules for adjusting or stopping the trial need to be determined {\it a priori} based on assumptions about potential effect sizes and the likely data that will be observed in the trial. 
If such assumptions are misspecified, then incorrect conclusions about the treatment effect can be made as a result of the design.\cite{thall_mueller_xu_guindani_2017} 
This presents significant challenges for adaptive trials as there can be considerable uncertainty about the onset, course and resolution of the disease. 
To address this challenge for trials with time-to-event outcomes, we propose a general Bayesian approach to trial design based on the partial likelihood that requires the data generating process to only be implicitly defined.\cite{bissiri_etal_2016}
We demonstrate the application of our approach via an illustrative example, and then apply it to redesign an ongoing clinical trial.

The focus of this paper is on Bayesian methods to design adaptive trials.
This was motivated by the inference framework providing a natural and principled approach to combine and update prior information as data are accrued, and enabling the quantification of evidence for or against a hypothesis.
The design of Bayesian adaptive trials is governed by the primary objective of the trial which is typically to draw conclusions about the effect of a treatment in a particular population. The true effect of the treatment is known as the {\it estimand}, and this is what will be estimated from the trial data.\cite{estimand_2021}
The rule for declaring the estimate as significant is defined from the outset, and is usually based on accruing sufficient evidence to achieve a certain benefit e.g.\ a posterior probability that the treatment effect is larger than a minimally important clinical difference.
Rules for adaptions at interim analyses are also defined, and typically relate to participant safety, efficiency in terms of use of trial resources and/or reducing burden on participants.
Once prior information about the potential treatment effect and the data generating process is ascertained, a large number of hypothetical trials are simulated to estimate trial metrics that provide insight into the performance of the design e.g.\ statistical power and Type I error. \cite{hummel_2015}
The trial decision rules, maximum sample size and timing and frequency of interim analyses can then be modified to optimise these metrics until an effective and efficient trial design is reached. 
The literature on Bayesian adaptive trial design is extensive, so we will not provide a comprehensive review in this paper. 
Instead, we direct the reader to a recent review and a seminal text.\cite{giovagnoli_2021,berry_book}

The use of trial simulation to evaluate Bayesian adaptive designs means that typically the design will depend on assumptions about the data that may be observed in the trial.\cite{berry_ho_1988,sylvester_1988,stallard_1998}
This could be an assumed treatment effect and/or a data generating process.
To address this limitation, we propose to consider general Bayesian methods where inference is performed based on a loss function which connects model parameters to the data but does not necessarily require specification of a probability distribution for the outcomes. 
Accordingly, such an inference method can provide robustness to a misspecified analysis model (i.e.\ where one fails to appropriately describe the underlying data generating process) and/or to potential contamination from outliers or anomalous data.  
Providing a framework for Bayesian learning in this way unified a variety of approaches that facilitate robust Bayesian inference.  
For example, Bayesian inference based on divergence metrics such as $\alpha$, $\beta$ and $\gamma$-divergence \cite{hooker_vid_2014,knoblauch_et_al_2018,ghosh_basu_2016} and the maximum mean discrepancy \cite{mmd_bayes_2020} can be couched within this framework. 
Further, generalised likelihood functions such as composite likelihoods,\cite{ribatet_cooley_davison_2012} quasi-likelihoods \cite{agnoletto_rigon_dunson_2023} and the empirical likelihood \cite{lazar_2003} can be considered within this framework via setting the loss function to be the negative generalised log-likelihood.  
Further, a variety of extensions to the general Bayesian approach have been considered for robust inference in different settings and include, for example, for Bayesian prediction via variational methods,\cite{frazier_2022} for generalised Bayesian filtering \cite{boustati_2020} and for Bayesian inference for models with intractable likelihoods.\cite{matsubara_2022}

The motivation for this research was the redesign of an ongoing clinical trial to assess the effectiveness of a third dose of a vaccine to protect against gastroenteritis in Australian Indigenous infants.  
This trial has a time-to-event outcome so we proposed to draw inference about the effect of the treatment via a general Bayesian posterior distribution where the loss function is based on the partial likelihood.\cite{cox_1975}  
The advantage of this Bayesian method is that the underlying baseline hazard function does not need to be defined in order to draw inference about the treatment effect.
Through leveraging this property, we show how to design trials for time-to-event outcomes where the data generating process is only implicitly defined.  
The main benefit of this proposed approach is that it largely removes concerns about model misspecification both at the design and analysis stages of the trial.
We believe this is the first paper to show how trial designs can be formed within a general Bayesian framework.

The remainder of the paper proceeds as follows.  In Section \ref{sec:mot_exp}, our motivating clinical trial is described.  
In Section \ref{sec:inference}, we provide background information on Bayesian inference for time-to-event outcomes.  
Bayesian design approaches are then described in Section \ref{std_design}.
These approaches are then extended in Section \ref{sec:general_des} to allow Bayesian design to be undertaken based on a general Bayesian posterior distribution.
In Section \ref{sec:illustrative}, an illustrative example is provided to demonstrate our methodology. 
Section \ref{sec:orvac} applies our methodology to redesign our motivating clinical trial.  
The paper concludes with Section \ref{sec:conc} where our approach is discussed along with some limitations and avenues for future research.

\section{Motivating clinical trial}\label{sec:mot_exp}

The Optimising Rotavirus Vaccine in Aboriginal Children (ORVAC) trial is a double-blind, randomised, placebo-controlled Bayesian adaptive clinical trial designed to test the effectiveness of a third dose of Rotarix rotavirus vaccine in Australian Indigenous infants in providing improved protection (versus usual care) against gastroenteritis.\cite{middleton_et_al_2019,jones_et_al_2020}
The third dose was considered as the active treatment with usual care considered as a matched placebo.
These two treatments were randomly allocated to participants at a ratio of 1:1, and this remained fixed for the entire trial.
There were two primary outcomes of the trial: (1) anti-rotavirus IgA seroconversion, defined as serum anti-rotavirus IgA $\geq$ 20 U/ml 28 to 55 days post Rotarix/placebo, and (2) time from randomisation to medical attendance for which the primary reason for presentation was acute gastroenteritis or acute diarrhoea illness.
For the purposes of this paper, we focus on the second primary outcome where the estimand is the hazard ratio between active groups and placebo.
To be eligible for enrolment in the trial, a participant needed to be aged between $6$ and $12$ months, and each participant was followed up to 36 months of age.
The maximum sample size for the trial was 1,000 participants, and planned analyses occurred after 250 participants had been enrolled initially, and then every 50 thereafter.  
At each planned analysis, pre-specified decision rules were evaluated that allowed the trial to stop due to sufficient evidence of treatment effectiveness or due to sufficient evidence to declare continuing the trial would be futile.


To design the original ORVAC trial, assumptions were made about the data generating process for the trial data.  
For the time-to-event outcome, a hazard function, which quantifies the instantaneous rate of the event at time since enrolment $t$, was defined, and the two treatment groups were assumed to have proportional hazards over time. \cite{cox_1972}
In such a case, the hazard function can be described as follows:
\begin{equation}
h(t \vert \bm{\theta}) = h(t \vert \beta, \bm{\tau}) = \exp (x \beta) h_0(t \vert \bm{\tau}),
\label{eqn:prophaz}
\end{equation}
where $x$ denotes the treatment allocation (e.g.\ $x = 1$ for active, and $x=0$ for placebo), $\beta \in \mathbb{R}$ denotes the treatment effect we wish to estimate, and $h_0(t\vert \bm{\tau})$ denotes the baseline hazard function, potentially depending on nuisance parameters $\bm{\tau} = \left(\tau_1, \dots, \tau_M\right)^T \in \mathcal{T}$. 
Lastly, $\bm{\theta} = (\beta,\bm{\tau}^T)^T\in\Theta$. 
Note that specification of the baseline hazard completely specifies the probability distribution of the outcomes. 
For example, if $h_0(t\vert \bm{\tau}) = \tau_1 \tau_2 t^{\tau_2-1}$, for $\tau_1>0$ and $\tau_2>0$, then the outcomes have a Weibull distribution.

Trial metrics for ORVAC were evaluated via simulation to assess the proposed design. 
This was undertaken based on specific forms of the baseline hazard function corresponding to the Weibull and exponential distributions. \cite{jones_et_al_2020}
Note that the exponential is a special case of the Weibull when $\tau_2=1$. 
Throughout this paper, to ensure comparisons can be made with the original design, the decision rules, the rule for concluding treatment effectiveness, maximum sample size, enrolment rate and timing and frequency of interim analyses will remain as specified within ORVAC.



\section{Inference framework}\label{sec:inference}

Throughout this paper, all inference will be performed within a Bayesian framework based on a posterior distribution.  
In standard Bayesian inference, a prior distribution is updated based on information from the data through the likelihood function.
For the ORVAC trial, each participant $i$ will yield outcome $t_i$ based on a treatment allocation denoted by $x_i$.
Then, assuming each observation is conditionally independent and subject to uninformative right censoring,\cite{cox_1983} the likelihood is:
\[
p(\bm{y}|\bm{\theta},\bm{x}) = \prod_{i=1}^N f(t_i|\bm{\theta},x_i)^{c_i} S(t_i|\bm{\theta},x_i)^{1-c_i},
\]
where $\bm{x}=(x_1,\ldots,x_N)^T$, $\bm{y} = (\bm{y}_1,\ldots,\bm{y}_N)^T$,  $\bm{y}_i = (t_i,c_i)$ such that $c_i = 0$ if the observation is censored and $c_i = 1$ otherwise and $f(.)$ and $S(.)$ denote the probability density and survivor functions (respectively) which follow from the specification of the hazard function in Equation \eqref{eqn:prophaz}, for $i=1,\ldots,N$.\cite{collett_2015}

With the above likelihood, the standard Bayesian (marginal) posterior distribution of the treatment effect $\beta$ can be defined as follows:
\[
p(\beta|\bm{y},\bm{x})\propto \int_{\mathcal{T}}p(\bm{y}|\bm{\theta},\bm{x}) p(\bm{\theta})d\bm{\tau},
\]
where $p(\bm{\theta})$ is the prior distribution on $\bm{\theta} = (\beta,\bm{\tau}^T)^T$.  
Given the above posterior distribution generally will not have a known form, Markov chain Monte Carlo (MCMC) methods can be used to approximate it.

Recently, general Bayesian inference has been proposed as a coherent procedure for updating prior beliefs based on a loss function rather than a likelihood function.
Consider a loss function denoted by $l(\beta, \bm{y}, \bm{x})$, which identifies desirable parameters of interest $\beta$ given $\bm{y}$ and $\bm{x}$. Then, the general Bayesian posterior distribution for $\beta$ is given by 
\begin{equation}\label{eq:general_bayes}
p(\beta|\bm{y},\bm{x}) \propto \exp\{-w l(\beta,\bm{y},\bm{x})\}p(\beta),
\end{equation}
where $w>0$ controls the rate of learning about $\beta$ from prior, $p(\beta)$, to posterior.  
Of note, standard Bayesian inference can be considered a special case of general Bayesian inference where the negative log-likelihood is the loss function (with $w=1$). 
The general Bayesian posterior provides a coherent update of prior information about the parameter values that minimise the expectation of the loss, $l(\beta, \bm{y}, \bm{x})$, with respect to the true data generating process.\cite{bissiri_etal_2016}



For the general Bayesian analysis of time-to-event outcomes, the negative partial log-likelihood \cite{cox_1975} has been considered as the loss function. \cite{bissiri_etal_2016,kalbeisch_1978, sinha_2003,  miller_2021}
For continuous time and distinct time-to-event outcomes, the loss function is
\begin{equation}\label{eq:partial}
l(\beta,\bm{y},\bm{x}) = -\sum_{i=1}^N c_i \log \frac{\exp(x_i\beta)}{\sum_{j\in R_i} \exp(x_j\beta)},
\end{equation}
where $R_i$ denotes the risk set (i.e.\ the set of participants who have not yet experienced the event of interest and are not censored) at the $i$th time $t_i$, for $i=1,\dots,N$. 
To calibrate the posterior based on this loss function, $w$ can be set to 1 such that properties of the general posterior match those of the maximum partial likelihood estimator.\cite{bissiri_etal_2016}

The above partial likelihood has often been used in medical research to estimate a treatment effect based on time-to-event outcomes.\cite{collett_2015}
The general posterior distribution based on this partial likelihood has also been studied previously.  
In particular, the marginal posterior distribution of $\beta$ (having integrated out the baseline hazard under a non-informative prior) converges to the general Bayesian posterior, formed via the partial likelihood, as $N \to \infty$.\cite{kalbeisch_1978}
In addition, under fairly mild sufficient conditions, the above posterior exhibits concentration, asymptotic normality and asymptotic frequentist coverage.\cite{miller_2021}



\section{Adaptive clinical trial design}\label{std_design}

Trial simulation is often used to evaluate Bayesian adaptive clinical trials.
This involves progressively simulating participant enrolment, the allocation of treatments and outcome data throughout the trial period, and evaluating decision rules at interim analyses. 
When the trial completes, either due to a stopping rule being triggered or enrolments reaching the maximum sample size, all trial data are analysed to address the question of interest.
Trial metrics such as statistical power and Type I error are then evaluated (typically based on the primary outcome) to provide insight into the expected 
performance of the trial design.
The trial design can then be adjusted to, in some way, optimise these trial metrics to yield an effective and efficient trial.



An approach to simulate a Bayesian adaptive clinical trial where time-to-event outcomes are observed is outlined in Algorithm \ref{alg:sim}.  
As an input to the algorithm, a data generating process $h(t|\beta,\bm{\tau})$ is assumed along with how the data will be analysed with associated prior information.  
In addition, other inputs such as the randomisation schedule, enrolment rate, timing of interim analyses and the decision rules are specified.  
Then, indexing trial time by $k$, which could represent a month, the trial is iteratively simulated for a total of $K$ time increments until the maximum sample size ($N_{\max}$) is reached or the trial is stopped as a result of a decision rule being triggered.  
At each $k$, $m_k\geq 1$ participants are enrolled (line 3), and each participant is randomly allocated a treatment according to the randomisation schedule (line 4).  
Baseline and other pre-treatment variables could also be simulated at this time.
However, these are not denoted within Algorithm \ref{alg:sim}.
Outcome data for these participants are then simulated based on their treatment allocation and the assumed data generating process (line 5).  
This continues iteratively until it is time for a planned interim analysis (line 6).  
When an interim analysis occurs, the simulated outcome data need to be considered in light of the time increment $k$.  
That is, if the first interim analysis occurs at month 3 and a participant has an event time of 6 months, then this observation cannot be considered (as is) within the interim analysis.  
Instead, it is appropriate to consider the observation as being right censored at 3 months (if the participant has been enrolled for the whole trial duration).  
Let the observations that have been censored due to the timing of an interim analysis be indexed by $r_k$ and denoted by $\bm{y}^*_{r_k}$ such that they are distinct from $\bm{y}_{r_k}$; the outcomes that would be observed, if the event was not censored due to the timing of the interim analysis.
Similarly, denote the responses from participants who have exited the trial (i.e.\ they have responded or are censored for a reason other than the timing of an interim analysis) as $\bm{y}_{-r_k}$.
Then, based on the data available at the interim analysis (i.e.\ $\bm{y}_{-r_k}$ and $\bm{y}^*_{r_k}$ and treatment allocations denoted by $\bm{x}_{1:n_k}$), a posterior distribution $p(\beta | \bm{y}_{-r_k},\bm{y}^*_{r_k},\bm{x}_{1:n_k})$ is computed, and the decision rules are evaluated based on this posterior (lines 7 to 8). 
If any decision rule is met, then the trial is stopped or adjusted, as appropriate (line 10).
Otherwise, the simulation continues iteratively until the maximum sample size is reached.
Once the trial has stopped or the maximum sample size has been reached, outcome data from all participants are analysed and conclusions about the effect of treatment are drawn (line 14 where $N$ denotes the total number of enrolments for the trial).

In the ORVAC trial, two decision rules were proposed to be evaluated at each interim analysis.  
The first rule was for declaring treatment effectiveness, and the second rule was for declaring that it would be futile to continue recruitment as there is a small chance of concluding treatment effectiveness if the maximum sample size was reached.  
Both of these decision rules were evaluated based on the primary outcome as described in Section \ref{sec:mot_exp} accounting for differences between treatments arms only.
We define both of these decision rules in the next section, and show how they can be evaluated at each interim analysis.

\subsection{Decision rule for stopping due to effectiveness}

For the ORVAC trial, effectiveness will be evaluated at interim analyses and also at the end of the trial once all participants have been followed-up.
Treatment effectiveness will be declared at the end of the trial if $P(\beta < 0 | \bm{y}_{1:N},\bm{x}_{1:N}) > \Delta$.
The rule for evaluating effectiveness at the interim analyses has the same form as this rule but is an expectation of the binary indicator (rather than the binary indicator itself) over unobserved outcomes, and thus we will refer to this as evaluating expected effectiveness.
To elaborate, consider that at the interim analyses there will be two types of participants: (1) participants who have been enrolled and have exited the trial (as they have either responded or their response has been censored);
or (2) participants who have been enrolled but have not exited the trial (as their time-to-event or censored outcome has not yet been observed).  
For the first type of participant, we have observed both their treatment allocation and their time-to-event outcome, so these are considered as they are when evaluating the decision rule.  
For the second type of participant, their treatment allocation is known and it is also known how long they have been enrolled in the trial without observing an event. 
Thus, there is uncertainty over their eventual time-to-event outcome, and this is the uncertainty the expectation of the effectiveness decision rule is evaluated over.
Specifically, this expectation can be defined as:
\[
E[\mathcal{I}(P(\beta < 0 | \bm{y}_{-r_k},\bm{Z},\bm{x}_{1:n_k}) > \Delta)] = \int_{\mathcal{Z}} \mathcal{I}(P(\beta < 0 | \bm{y}_{-r_k},\bm{z},\bm{x}_{1:n_k}) > \Delta)p(\bm{z}|\bm{y}_{-r_k},\bm{y}^*_{r_k},\bm{x}_{1:n_k})d\bm{z},
\]

where $\mathcal{I}()$ is an indicator function which equals one if the event is true and zero otherwise, and $\bm{Z} \in \mathcal{Z}$ is the random variable associated with $\bm{z}$ which denotes supposed future realisations of the time-to-event for the second type of participant. 
Note that, for ease of notation, we have (for the moment) ignored any censoring that may be associated with $\bm{z}$.

The marginal distribution of the random variable $\bm{Z}$ can be defined as follows:

\begin{equation}\label{eq:pz}
    p(\bm{z}|\bm{y}_{-r_k},\bm{y}^*_{r_k},\bm{x}_{1:n_k}) = \int_{\Theta} p(\bm{z}|\bm{\theta},\bm{y}^*_{r_k},\bm{x}_{r_k})p(\bm{\theta} | \bm{y}_{-r_k},\bm{y}^*_{r_k},\bm{x}_{1:n_k})d\bm{\theta},
\end{equation}

where 
\[
p(\bm{z}|\bm{\theta},\bm{y}^*_{r_k},\bm{x}_{r_k}) = \frac{p(\bm{z} | \bm{\theta}, \bm{x}_{r_k})}{1-F_Z(\bm{y}^*_{r_k}| \bm{\theta}, \bm{x}_{r_k})}\mathcal{I}(\bm{z}>\bm{y}^*_{r_k}),\]

$p(\bm{\theta} | \bm{y}_{-r_k},\bm{y}^*_{r_k},\bm{x}_{1:n_k})$ denotes the posterior distribution of $\bm{\theta}$ based on the data available at an interim analysis, $p(\bm{z} | \bm{\theta}, \bm{x}_{r_k})$ is the likelihood function, and $F_Z(\bm{y}^*_{r_k}| \bm{\theta}, \bm{x}_{r_k})$ denotes the cumulative distribution function for $Z$ evaluated at $\bm{y}^*_{r_k}$.
Thus, the marginal distribution of $\bm{z}$ is the posterior predictive distribution truncated such that $\bm{z}>\bm{y}^*_{r_k}$.

In practice, evaluating the expectation of the effectiveness decision rule is analytically intractable so Monte Carlo methods can be used to form an estimate as follows:
\begin{equation}\label{eq:suc}
E[\mathcal{I}(P(\beta < 0 | \bm{y}_{-r_k},\bm{Z},\bm{x}_{1:n_k}) > \Delta)] \approx 1/B\sum_{b=1}^B \mathcal{I}(P(\beta < 0 | \bm{y}_{-r_k},\bm{z}_b,\bm{x}_{1:n_k}) > \Delta),
\end{equation}
where $\bm{z}_b \sim p(\bm{z}|\bm{y}_{-r_k},\bm{y}^*_{r_k},\bm{x}_{1:n_k})$.
If this expectation is sufficiently large then we would recommend the trial stop due to expected effectiveness.
Let $\Delta_e$ denote the threshold for this comparison for declaring expected effectiveness. 
As the notation suggests, can be distinct from $\Delta$; the threshold for declaring treatment effectiveness at the end of the trial once all participants have completed follow-up.

When evaluating the above approximation, appropriate censoring will need to be applied to $\bm{z}_b$.
For example, the simulated observations may need to be right-censored if they extend beyond the time remaining in the trial.
An additional consideration for the ORVAC trial is that participants are followed up until 36 months of age.
If they have not responded and are not censored by this age, then their outcome is right-censored.
This type of censoring would also need to be applied to these simulated outcomes, as appropriate.
Further details about this are given in Section \ref{sec:orvac}.


The approach to approximate expected effectiveness rule at each interim analysis is outlined in Algorithm \ref{alg:1} where the data $\bm{y}_{-r_k}$ and $\bm{y}^*_{r_k}$ and the treatment allocations $\bm{x}_{1:n_k}$ are inputs.  
For those participants who have not yet responded and are not censored, their treatment allocation is known but their outcome is not, so their outcome is simulated (line 3).
This involves simulating $\bm{z}$ from the marginal distribution defined in Equation \eqref{eq:pz} as follows: $\bm{z}_b \sim p(\bm{z}|\bm{\theta}_b,\bm{y}^*_{r_k},\bm{x}_{r_k})$ and $\bm{\theta}_b \sim p(\bm{\theta}|\bm{y}_{-r_k},\bm{y}^*_{r_k},\bm{x}_{1:n_k})$
which will yield left-truncated survival data.
As appropriate, these simulated outcomes may need to be censored as discussed above.
This forms a partly simulated data set which contains data for all participants enrolled in the trial.
Given this data set, an updated posterior distribution of the treatment effect (i.e. $p(\beta | \bm{y}_{-r_k},\bm{z}_b,\bm{x}_{1:n_k})$) is found (line 4).  
Based on this posterior, an indicator function is evaluated to determine whether trial effectiveness would be concluded at the pre-specified level of evidence $\Delta$, if the simulated outcomes from the remaining enrolled participants were ascertained (line 5).  
After repeating this procedure $B$ times, the expectation is approximated (line 7).  
If this is larger than a threshold $\Delta_e$, then the trial will stop due to expected effectiveness (lines 8 to 10).


\subsection{Decision rule for stopping due to futility}

The futility decision rule is used to determine whether it would be futile to continue the trial as it is unlikely that treatment effectiveness would be concluded if the maximum sample size was reached.  
This can be evaluated similarly to the effectiveness rule but now incorporates uncertainty about the outcomes from participants who have not yet enrolled into the trial.  
Again, to elaborate, consider that at each interim analysis there will be three types of participants: (1) and (2) from the above effectiveness decision rule; and (3) participants who have not yet enrolled and therefore do not have a treatment allocation, have not yet responded and are not censored.  
Thus, the expectation of the futility decision rule is evaluated over the uncertainty in the eventual responses from the second type of participant, and the uncertainty in treatment allocations and responses from the third type of participant.
This expectation can be defined as follows:
\begin{align*}
   E&[\mathcal{I}(P(\beta < 0 | \bm{y}_{-r_k},\bm{Z},\bm{W},\bm{x}_{1:n_k},\bm{V}) > \Delta)] = \\
&\int_{\mathcal{Z}}\int_{\mathcal{W}} \sum_{\bm{v}\in\mathcal{V}} \mathcal{I}(P(\beta<0|\bm{y}_{-r_k},\bm{z},\bm{w},\bm{x}_{1:n_k},\bm{v}) > \Delta)
p(\bm{w}|\bm{y}_{-r_k},\bm{y}^*_{r_k},\bm{x}_{1:n_k},\bm{v})
p(\bm{z}|\bm{y}_{-r_k},\bm{y}^*_{r_k},\bm{x}_{1:n_k})
p(\bm{v})
d\bm{w}d\bm{z},
\end{align*}

where $\bm{V}$ and $\bm{W}$ are the random variables for future treatment allocations and the associated future outcome data for the third type of participant, respectively, and we (for the moment) ignore any censoring associated with $\bm{z}$ and $\bm{w}$.  
The distribution of $\bm{V}$, denoted as $p(\bm{v})$, is based on the randomisation scheme for the trial.
Given the treatment allocations $\bm{v}$, the distribution of $\bm{W}$ can be defined as:
\begin{equation}\label{eq:w}
p(\bm{w}|\bm{y}_{-r_k},\bm{y}^*_{r_k},\bm{x}_{1:n_k},\bm{v}) = \int_{\Theta} p(\bm{w}|\bm{\theta},\bm{v})p(\bm{\theta} | \bm{y}_{-r_k},\bm{y}^*_{r_k},\bm{x}_{1:n_k})d\bm{\theta},\end{equation}
where $p(\bm{w}|\bm{\theta},\bm{v})$ is the likelihood function.
Thus, $\bm{w}$ are posterior predictive data (given $\bm{v}$).
If the above expectation is less than $\Delta_f$, then the trial would stop for futility.


The expectation of the futility decision rule is analytically intractable, so Monte Carlo methods can be used to approximate it as follows:
\begin{equation}\label{eq:fut}
E[\mathcal{I}(P(\beta < 0 | \bm{y}_{-r_k},\bm{Z},\bm{W},\bm{x}_{1:n_k},\bm{V}) > \Delta)] \approx 1/B\sum_{b=1}^B \mathcal{I}(P(\beta<0|\bm{y}_{-r_k},\bm{z}_b,\bm{w}_b,\bm{x}_{1:n_k},\bm{v}_b) > \Delta),
\end{equation}
where $\bm{w}_b\sim p(\bm{w}|\bm{y}_{-r_k},\bm{y}^*_{r_k},\bm{x}_{1:n_k},\bm{v}_b)$, $\bm{v}_b \sim p(\bm{v})$ and $\bm{z}_b$ is as defined above.
Outcomes $\bm{z}_b$ and $\bm{w}_b$ would need to be censored, as appropriate.

The approach to evaluate this approximation is similar to what was given above for the effectiveness decision rule, so the algorithm for this is in the Supplementary Material.
The main difference is that treatment allocations are simulated for participants who are not yet enrolled in the trial, and data for these participants are also generated, see Algorithm SM.1.

\section{Adaptive clinical trial design based on a general Bayesian posterior distribution}\label{sec:general_des}

In this section, we describe our approach to design Bayesian adaptive trials with time-to-event outcomes where the data generating process is only implicitly defined.
To facilitate this, we consider trial design within a general Bayesian inference framework.
A major challenge in doing so is that the loss function need not be completely linked to the sampling distribution of the data.
Specifically, here we cannot generate data from the partial likelihood defined in Equation (\ref{eq:partial}).
This means it is unclear how Algorithms \ref{alg:sim}, \ref{alg:1} and SM.1 can be used to evaluate trial designs.  
To overcome this, we propose to consider a {\it super model} which we specify such that it can generate a wide range of data sets that are potentially observable in the trial.
In this way, its purpose is to be flexible such that it can represent a wide range of potential data generating processes.
For the ORVAC trial, we propose to instill this flexibility in the baseline hazard function.
Formally, let $s(t|\beta,\bm{\psi}) = \exp(x\beta)s_0(t|\bm{\psi})$ denote the hazard function defined by the super model, where $\bm{\psi} = \left(\psi_1, \dots, \psi_Q\right)^T \in \mathcal{P}$.
To provide flexibility in the baseline hazard, we define $s_0(t|\bm{\psi})$ via a cubic spline representation as follows\cite{herndon_harrell_1995,harden_kropko_2019}:
\[
s_0(t \vert \bm{\psi}) = \sum_{q=1}^Q \psi_q g_q(t),
\]
where $g_q(t)=t^q$ are the basis functions, for $q=1,\ldots,Q$.

To complete the definition of the super model, a prior is placed on $Q$ and $\bm{\psi}$. 
Depending on this prior, the baseline hazard could capture a variety of forms and therefore a variety of data generating processes including those that are constant, monotonic or non-monotonic.
Of note, the super model will likely be an overparameterised descriptor of the trial data, and is therefore out of the scope of inference.
That is, it is not desirable to estimate a treatment effect based on the super model.  
Indeed, the super model could contain features or parameters which would be considered nuisance.

To undertake trial design based on general Bayesian methods, an approach to compute the general Bayesian posterior distribution (Equation \ref{eq:general_bayes}) is needed.
MCMC methods could be implemented, however, here the general Bayesian posterior consists of a single parameter only i.e.\ the treatment effect, so numerical integration could be adopted. \cite{ohagan_1991} 
When the general posterior distribution consists of additional parameters, approximate methods could be implemented for computational efficiency (i.e.\ across a large number of simulations).
This could be, for example, a Laplace approximation which has the following mean:
\[
\beta^* = \arg\max_{\beta\in\mathcal{B}} \log p(\beta) - wl(\beta,\bm{y},\bm{x}),
\]
and a variance-covariance matrix that is the inverse of the negative Hessian matrix evaluated at $\beta^*$, where $l(\beta,\bm{y},\bm{x})$ is the negative partial log-likelihood as given in Equation (\ref{eq:partial}).  

Thus, we propose to undertake general Bayesian adaptive clinical trial design by replacing the data generating process with a super model, and performing all inference based on the general posterior distribution.  
We consider this approach next via an illustrative example and for the redesign of the ORVAC trial.

\section{Illustrative example}\label{sec:illustrative}

Before re-designing the ORVAC trial, we consider an illustrative example to demonstrate our methodology.
Suppose two treatments (active and placebo) are being compared based on a time-to-event outcome, and that an appropriate sample size is required to maintain at least 80\% power to declare the active treatment as being preferred over placebo based on the results of a Bayesian analysis.
Suppose the duration of the trial is 40 months, all participants will be enrolled at the start of the trial and any participant who has not experienced their outcome by the end of the trial is right censored.

\begin{figure}[h!]
\centering
\includegraphics[width=1.0\textwidth]{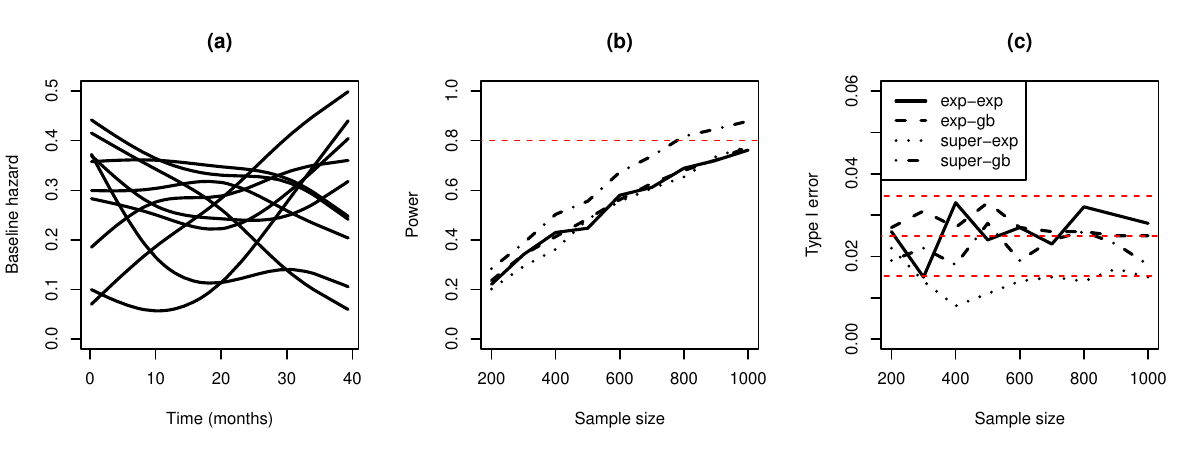}
\caption{\label{fig:toy_power_type1}(a) Ten realisations of the baseline hazard function when defined via the super model. Simulation results (i.e.\ (b) Power and (c) Type I error) when the exponential (`exp') and super models generated the data and the exponential and general Bayesian (`gb') models were used to estimate the treatment effect. All results for power and Type I error are based on 1,000 trial simulations.}
\end{figure}

To determine an appropriate sample size $N$ (i.e.\ the design parameter), a treatment effect of $\beta=-0.2$ was assumed, and power was evaluated via simulation based on the rule $P(\beta < 0|\bm{y}_{1:N},\bm{x}_{1:N}) > 0.975$.
That is, if the posterior probability of the treatment effect being less than 0 was greater than 0.975, then it was declared that active treatment was preferred over placebo.
To demonstrate our approach and compare it with a standard Bayesian method, four scenarios were considered.
These scenarios differed in terms of assumptions about the data generating process, and the analysis model used to form a posterior distribution for the treatment effect (and thus to evaluate power). 
Specifically, two models were considered for data generation.
The first was the exponential proportional hazards model where a uniform prior was placed on $\log\tau_1$ with bounds $-3.72$ and $-2.72$.
The second data generating model was a super model as defined above based on a cubic spline representation of the baseline hazard function.
A point prior was placed on the number of basis functions i.e.\ $Q=3$, and a prior was induced on $\bm{\psi}$ by assuming values of the baseline hazard were uniformly distributed between 0 and 0.5 at times (or `knots') 0, 10, 20, 30 and 40, which represent the trial time in months.
The range of values for $\bm{\psi}$ that could be observed based on this prior yielded a distribution for $\bm{\psi}$ which is thus the induced prior.
Figure \ref{fig:toy_power_type1}(a) shows 10 realisations of the baseline hazard function when the super model was defined in this way.
As can be seen, a variety of forms can be observed including some that are relatively flat, monotonic and non-monotonic in nature.

For each of the above data generating models, two approaches were used to form a posterior distribution for the treatment effect.
One approach was standard Bayesian inference based on the exponential proportional hazards model and the other was general Bayesian inference (Equation \ref{eq:general_bayes}) via the partial likelihood (Equation \ref{eq:partial}).
In both cases, vaguely informative prior information was assumed.
In addition to power, Type 1 error was also evaluated under the null scenario i.e.\ when $\beta=0$.

Figure \ref{fig:toy_power_type1}(b) shows the results for the power analysis based on 1,000 simulations of the four scenarios described above.
When the exponential model was used for both data generation and as the analysis model, power approached 80\% when the sample size approached 1,000.
When the exponential model generated the data and the general Bayesian model was the analysis model, a similar power profile was observed.
This means no reduction in power was observed when the treatment effect was estimated via an undefined baseline hazard function compared to when this function was correctly specified.
When the super model generated the data and the general Bayesian model was the analysis model, relatively large values of power were observed i.e.\ power reached 80\% when the sample size was 800.
Of interest was when the super model generated the data and the analysis model was the exponential model.
Here, there was a reduction in power across most values for $N$, with power again only approaching 80\% when the sample size was 1,000.
This shows the potential reduction in power that can occur when the analysis model is inflexible to departures from the assumed baseline hazard function.

The Type I error was also evaluated based on 1,000 simulations of the four scenarios described above, see Figure \ref{fig:toy_power_type1}(c).
Given the rule for declaring treatment effectiveness, the Type I error was expected to be 0.025 (denoted by the middle red dashed line, with the other two red dashed lines denoting the margin of error).
As can be seen, the Type I error was generally around 0.025 and within the margin of error for three of the four scenarios.
The only scenario where this was not observed was when the super model generated the data and the exponential model was the analysis model.
In this case, generally lower than expected Type I errors were observed.
Investigating this further revealed that this may be due to the exponential model underestimating the effect of treatment, which may also explain the reduced power in this scenario.


Overall, the results from this illustrative example suggest that selecting a sample size based on a general Bayesian approach is preferred over a standard approach.
This is because the general Bayesian approach yielded a lower sample size to achieve 80\% power, and maintained an appropriate Type I error.

\section{Redesigning the ORVAC trial}\label{sec:orvac}

We consider redesigning the ORVAC trial as described in Section \ref{sec:mot_exp}, and evaluated designs through simulation as shown in Algorithm \ref{alg:sim}, with expected effectiveness and futility being evaluated at interim analyses as shown in Algorithms \ref{alg:1} and SM.1, respectively.
For all simulations, the treatment was deemed effective at the end of the trial if $P(\beta<0 | \bm{y}_{1:N_{j}}, \bm{x}_{1:N_{j}})>0.97$ i.e. $\Delta=0.97$, where $N_{j}$ denotes the total number of enrolments in the $j$th simulation, for $j=1,\ldots,J$.\cite{jones_et_al_2020}
If $N_{j} < N_{\max}$ then the $j$th simulation stopped early either due to expected effectiveness or futility (where $N_{\max} = 1000$).
The thresholds for declaring expected effectiveness or futility at interim analyses (i.e.\ $\Delta_e$ and $\Delta_f$) were set to 0.90 and 0.05, respectively.\cite{jones_et_al_2020}
 
Throughout the simulations, data were generated from three different models.
Two of these models were considered when originally designing the trial, and were the exponential and Weibull proportional hazards models, as defined in Section \ref{sec:mot_exp}.  
The third model was the super model as defined in the illustrative example.
To draw inference about the treatment effect, four different analysis models were considered.
The first two analysis models were the exponential and Weibull proportional hazards models as originally implemented.\cite{jones_et_al_2020}
Both of these models were considered within a standard Bayesian inference framework.
For the third model, the treatment effect was estimated via the general Bayesian model as described in the illustrative example i.e.\ via Equation (\ref{eq:general_bayes}) with the negative partial log-likelihood as the loss function.
For the fourth model, a Bayesian semiparametric approach was adopted via the piecewise exponential hazards model.\cite{sinha_dey_1997}
For each simulated data set, to determine the appropriate number of `pieces', a reversible jump MCMC scheme was adopted as implemented via the R package `BayesReversePLLH'. \cite{chapple_package}
Thus, there were twelve scenarios made up of three different models for data generation and four different analysis models.
Of interest was comparing estimates of the treatment effect, the proportion of trials where treatment effectiveness or futility was declared, and assessing differences in the conduct of the trial e.g.\ whether certain trials stopped earlier than others.

To evaluate the decision rules for ORVAC, as shown in Algorithms \ref{alg:1} and SM.1, data needed to be generated from the marginal distributions of $\bm{z}$ and $\bm{w}$, see Equations \eqref{eq:pz} and \eqref{eq:w}.
In particular, these marginal distributions needed to be updated at interim analyses such that the resulting posterior predictive data were generally more aligned with the trial data.
To determine these marginal distributions, the posterior distribution of $\bm{\theta}$ was needed i.e.\ $p(\bm{\theta}|\bm{y}_{-r_k},\bm{y}^*_{r_k},\bm{x}_{1:n_k})$.
For the exponential and Weibull models, finding the posterior distribution was rather straightforward i.e.\ standard Bayesian inference.  
For the super model, finding the posterior distribution was also undertaken within a standard Bayesian framework but requires some explanation. 
To update the super model, we fixed $Q$ and the knot positions as described in the illustrative example.  
The values corresponding to the knots were then treated as free parameters to be estimated based on a given data set.  
That is, given a value of these free parameters, values of the hazard function were evaluated based on the simulated data, and these values were then converted to likelihood values.  
Standard Bayesian estimation then proceeded to determine the posterior distribution of $\bm{\theta}$.

Other features of this trial that needed to be considered are that, for a participant to be eligible for enrolment, they must be aged between 6 and $12$ months old, and they are followed up until they are 36 months old.
If they have not responded and are not censored by this time, then their event time is how long they have been enrolled in the trial and is right-censored.
To simulate this within the trial, we assumed 50 participants would be enrolled every three months with an age that was randomly drawn from a uniform distribution within bounds at 6 and 12 months.  
Then the age of each participant was tracked through each simulation i.e.\ increased at every iteration of the simulation so that they could be followed-up for an appropriate duration of time, and also appropriately right-censored when evaluating the decision rules.  
Age was not considered as a covariate in any model.

All results discussed below are based on 500 simulations.
We initially assumed each of the three data generating models had specific baseline hazard functions i.e.\ each model had a point prior.
With this assumption, we explored the behaviour of the general Bayesian posterior as the sample size increased.
This was then extended to assess estimation of the treatment effect and different trial metrics under the four analysis models within a fixed sample size simulation (i.e.\ without interim analyses) and then within an adaptive trial simulation (i.e.\ with interim analyses).
The adaptive trial simulation was then extended to consider a range of baseline hazards functions which could be observed under the super model.


\subsection{Behaviour of the general posterior distribution with increasing $N$}

\begin{figure}[h!]
\centering
\includegraphics[width=1.0\textwidth]{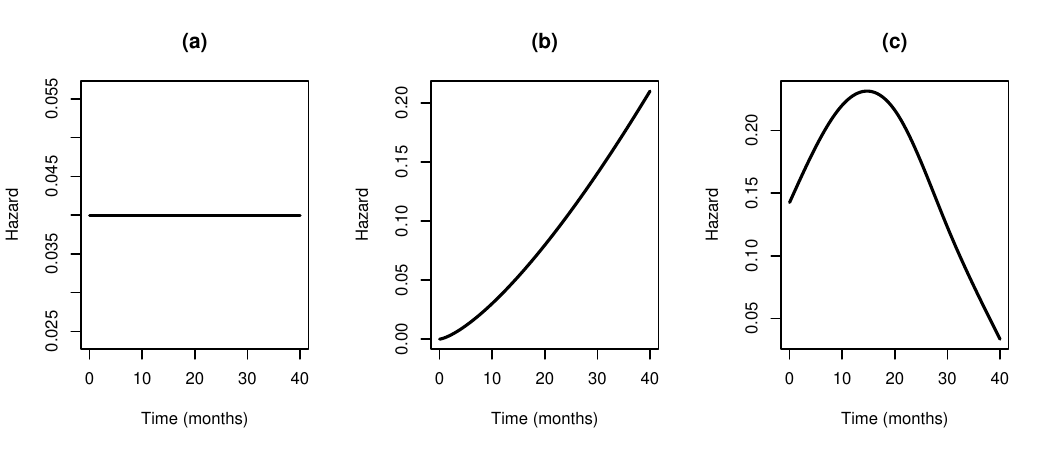}
\caption{\label{fig:plot_hazards}Specific baseline hazard functions under the (a) exponential, (b) Weibull and (c) super models.}
\end{figure}

Initially we explored the behaviour of the general posterior distribution based on the partial likelihood as the sample size increased.
To do so, we considered trial simulation based on a variety of different sample sizes where three models (as described above) were used to generate data i.e.\ the exponential, Weibull and the super model.
Under each model, specific forms of the hazard function were assumed, and these are shown in Figure \ref{fig:plot_hazards}.  
Treatment allocations within each simulation were randomly assigned 1:1, and a variety of treatment effects were assumed (i.e.\ $\beta\in \{0, -0.075, -0.125, -0.175, -0.25, -0.5\}$).  
Once data were generated, the general posterior distribution of the treatment effect was formed, and the posterior mode was recorded.  
This was repeated 1,000 times for each sample size.

\begin{figure}[h!]
\centering
\includegraphics[width=1.0\textwidth]{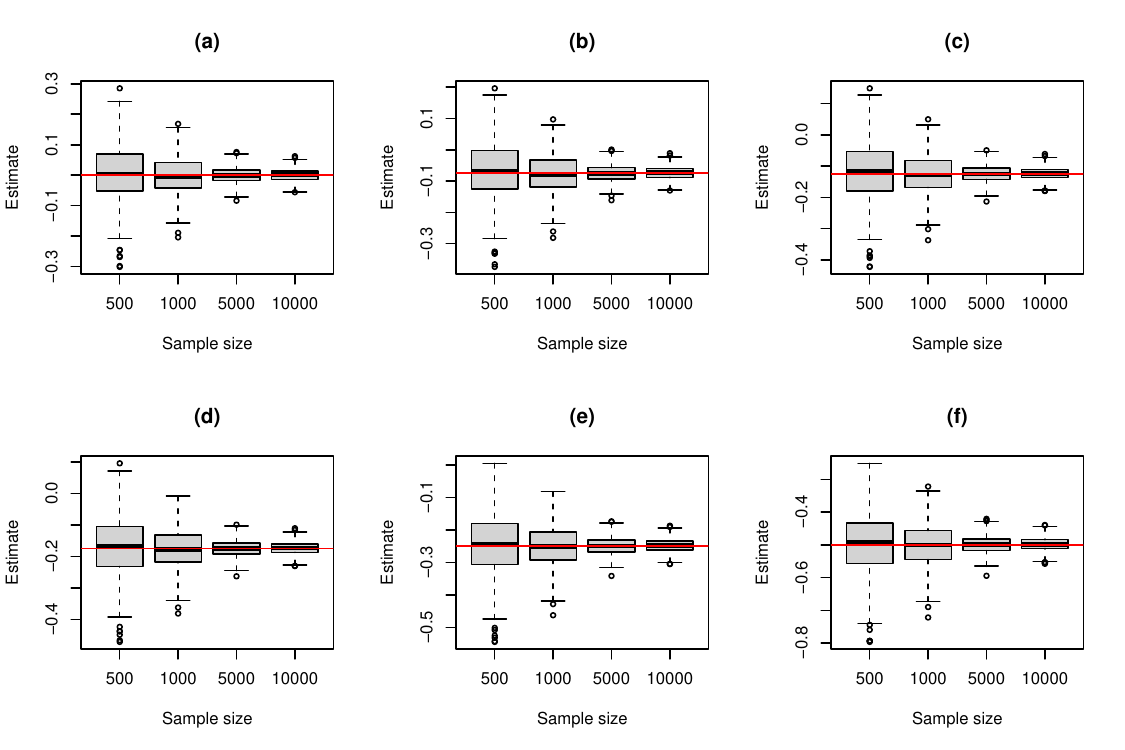}
\caption{\label{fig:large_n_super}Distribution of posterior modes for the treatment effect under different sample sizes when data were generated from the super model with (a) $\beta=0$, (b) $\beta=-0.075$, (c) $\beta=-0.125$, (d) $\beta=-0.175$, (e) $\beta=-0.25$, (f) $\beta=-0.5$ and the treatment effect was estimated based on the general posterior. All results are based on 500 trial simulations.}
\end{figure}

The results from this simulation are shown in Figure \ref{fig:large_n_super} for when the super model generated data, with results for when the exponential and Weibull model generated the data shown in the Supplementary Material, see Figures SM.2 to SM.3, respectively.
Each of these figures show the distribution of the posterior modes for the treatment effect for different sample sizes.
As can be seen, the central tendency of the posterior modes were around the assumed value, with any discrepancies only being minor and for the smallest sample size considered.
This suggests the approach is unbiased.
In addition, as the sample size increased, the variance of the posterior modes consistently decreased so all modes were progressively becoming more closely distributed around the assumed value.
This suggests the general Bayesian approach exhibits concentration which agrees with theoretical results. \cite{miller_2021}

\subsection{Fixed sample size trial simulation under specific hazard functions}

\begin{figure}[h!]
\centering
\includegraphics[width=1.0\textwidth]{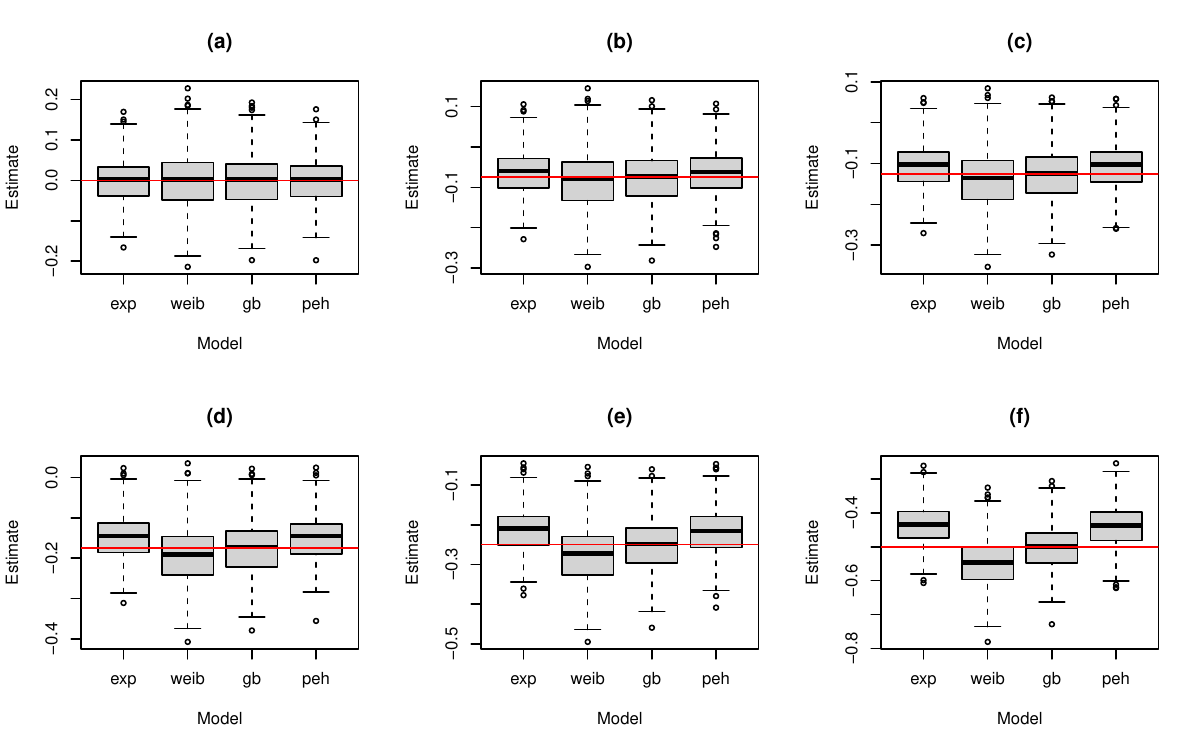}
\caption{\label{fig:sim_est_super}Distribution of posterior modes for the treatment effect from trial simulation with a fixed sample size when data were generated from the super model with (a) $\beta=0$, (b) $\beta=-0.075$, (c) $\beta=-0.125$, (d) $\beta=-0.175$, (e) $\beta=-0.25$, (f) $\beta=-0.5$ and the treatment effect was estimated based on the exponential (`exp'), Weibull (`weib'), general Bayesian (`gb') and piecewise exponential hazard (`peh') models. All results are based on 500 trial simulations.}
\end{figure}

The above simulation study was extended to consider all four analysis models (as described above) with a fixed sample size of 1,000.
The results from this simulation study are shown in Figure \ref{fig:sim_est_super} which shows the distribution of the posterior modes when the super model generated the data. 
The corresponding plots for when the exponential and Weibull models generated the data are shown in the Supplementary Material, see Figures SM.4 and SM.5, respectively.  
What is clear from these distributions is that biased estimates of the treatment effect were observed depending on the model used to generate data and the model used to draw inference.
For example, when the exponential model generated the data, all models provided relatively unbiased estimates of treatment effect with similar spread around the mean.
When the Weibull model generated data, the exponential model underestimated the treatment effect while the Weibull and general Bayesian models provided relatively unbiased estimates.
The piecewise exponential model yielded relatively unbiased estimates for small to moderate effect sizes but biased estimates were observed for larger effect sizes.  
When the data were generated from the super model, the exponential and piecewise exponential models generally underestimated the effect of treatment, while the Weibull model overestimated this effect. 
This under/overestimation became worse as the effect of treatment increased.  

The above results are reflected in the mean squared errors (MSEs) given in Table \ref{tab:mse_trial} which compare the posterior modes to the assumed treatment effect.
In particular, all analysis models provided similar MSEs when the exponential model generated the data, and the Weibull and general Bayesian models provided similar MSEs when the Weibull model generated the data.
When the super model generated the data, the MSEs for the general Bayesian model were consistently low while the MSEs for the other three analysis models had substantial increases with the assumed treatment effect.

Table \ref{tab:type11} shows the proportion of times active treatment was declared effective compared to placebo (i.e.\ $P(\beta <0 |\bm{y}_{1:N_{\max}},\bm{x}_{1:N_{\max}}) < \Delta$) for all combinations of three data generating models and four analysis models.
When the treatment had a non-zero effect, this can be interpreted as statistical power.
Under the null treatment effect, this can be interpreted as the Type I error. 
Given the rule for declaring treatment effectiveness, we would expect the proportion of Type I errors to be approximately 0.03.
As can be seen, when the exponential model generated the data, similar power and Type I errors were observed across the four analysis models.
When the Weibull model generated the data, both the Weibull and general Bayesian model yielded similar power and Type I errors while both the exponential and piecewise exponential models yielded reduced power and fewer Type I errors.
When the super model generated the data, the general Bayesian model yielded the most appropriate Type 1 error followed by the Weibull model while the exponential and peicewise exponential models yielded lower than expected Type I errors.
In terms of power, the Weibull model yielded slightly higher power than the general Bayesian model (which may align with the slightly inflated proportion of Type I errors).
The remaining two models yielded lower power overall.

Overall, regardless of which model generated the data, the general Bayesian model provided the least biased estimates of the treatment effect with similar variability to the alternative analysis models.
It was also the analysis model that maintained an appropriate proportion of Type I errors across all scenarios. 
In addition, power results under this analysis model were similar to the exponential and Weibull models when each of these models generated the data.

\subsection{Adaptive trial simulation under specific hazard functions}

Next, we extended the above simulation study to consider the adaptive design of the ORVAC trial.
Similar to the above approach, three models were considered for data generation, and the specific baseline hazard functions under each model are given in Figure \ref{fig:plot_hazards}.
Again, four analysis models were used to estimate the treatment effect and as the basis for evaluating the decision rules.  
From each simulation, we recorded $P(\beta < 0 |\bm{y}_{1:N_j},\bm{x}_{1:N_j}) > \Delta$.
In addition, we recorded the expectation of the effectiveness (Equation \ref{eq:suc}) and futility (Equation \ref{eq:fut}) rules based on the last interim analysis conducted in the $j$th simulation.
These summaries were recorded regardless of whether the maximum sample size was reached or the trial stopped early.

We start by inspecting the distribution of posterior modes shown in Figure \ref{fig:plot_mu0_v2} for when the super model generated the data, and shown in the Supplementary Material for when the exponential and Weibull models generated the data, see Figures SM.6 and SM.7, respectively.  
As can be seen, when data were generated from the exponential model, the posterior modes from all four analysis models were relatively unbiased, regardless of the assumed treatment effect.
When the Weibull model generated the data, the exponential model again underestimated the effect of treatment while the remaining three analysis models provided relatively unbiased estimates of this effect.
When the data were generated from the super model, again the exponential and piecewise exponential models underestimated the effect of treatment while the Weibull model overestimated this effect for the relatively moderate to large treatment effects.
Again, the general Bayesian model provided the least biased estimates of treatment effect overall, and this was regardless of the assumed value for $\beta$.

In terms of MSE, again all analysis models provided similar values when the exponential model generated the data, and the Weibull and general Bayesian model provided similar MSE values when the Weibull model generated the data, see Table \ref{tab:mse_adaptive_trial}.
When the super model generated the data, despite the exponential and piecewise exponential models providing biased estimates of treatment effect, these two models yielded the lowest MSE values.
This was due to the variance of the posterior modes being relatively low around an, on average, biased estimate.
Of the two least biased analysis models i.e.\ the Weibull and general Bayesian models, the general Bayesian model provided the lower MSE values overall.

\begin{figure}[h!]
\centering
\includegraphics[width=1.0\textwidth]{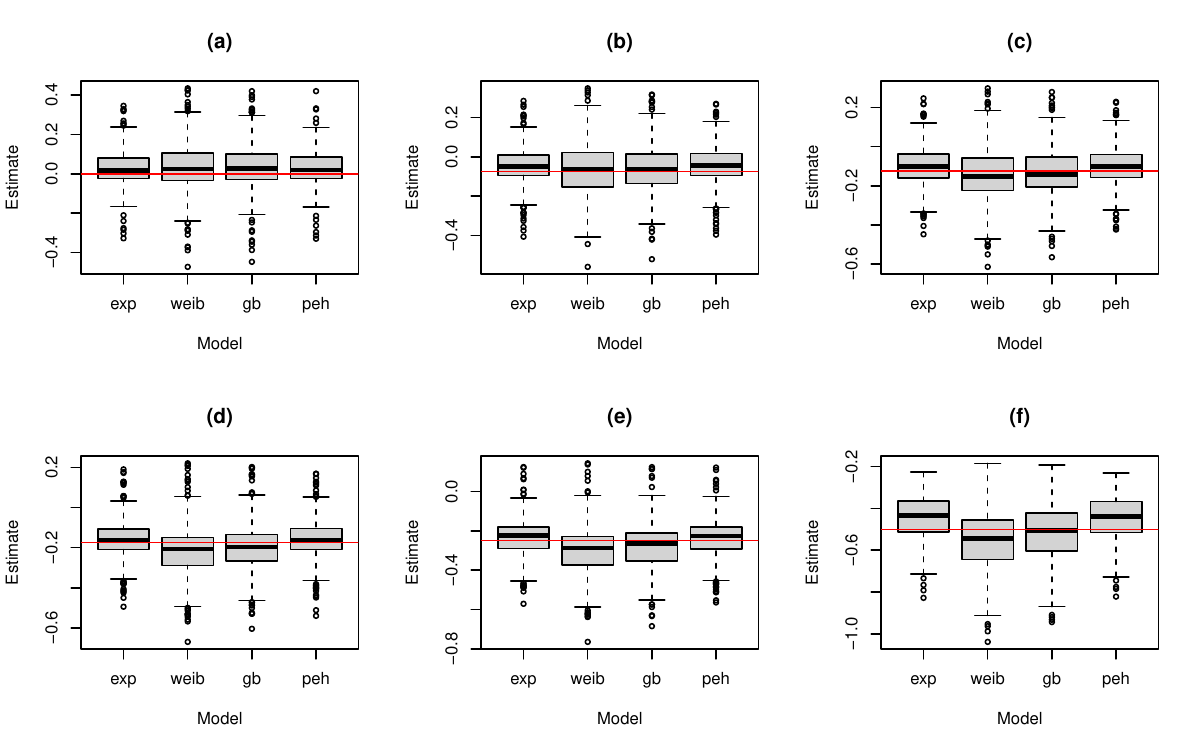}
\caption{\label{fig:plot_mu0_v2}Distribution of posterior modes for the treatment effect from trial simulation with an adaptive design when data were generated from the super model with (a) $\beta=0$, (b) $\beta=-0.075$, (c) $\beta=-0.125$, (d) $\beta=-0.175$, (e) $\beta=-0.25$, (f) $\beta=-0.5$ and the treatment effect was estimated based on the exponential (`exp'), Weibull (`weib'), general Bayesian (`gb') and piecewise exponential hazard (`peh') models. All results are based on 500 trial simulations.}
\end{figure}

Table \ref{tab:type12} shows the proportion of simulated trials where the active treatment was declared effective for all combinations of three data generating models and four analysis models.
As above, when $\beta < 0$, we can interpret this as power, and when $\beta=0$, we can interpret this as the Type I error.
Of note, given the simulated trials were able to stop early due to effectiveness or futility, it is not clear what proportion of Type I errors is expected.
For example, the inclusion of stopping rules can often inflate the chance of a Type I error.
Instead, when designing a trial, we would need to ensure this proportion remained low, in general e.g.\ around 0.03 when comparing an active treatment to placebo.
As can be seen, this proportion is low for the majority of scenarios, with the general Bayesian model yielding similar proportions to the exponential and Weibull models when each of these models generated the data.
When the super model generated the data, the proportion of Type I errors under the general Bayesian model remained reasonably low at 0.052, slightly higher for the Weibull model at 0.066 and lower under the exponential (0.024) and piecewise (0.020) models.
In terms of power, similar patterns were observed under the general Bayesian model compared to the exponential and Weibull models when each of these models generated the data, with perhaps some slight variation when the treatment only had a small effect.
When the super model generated the data, again the Weibull model yielded larger estimates of power than the general Bayesian model, while the other two models provided the two lowest estimates of power.

\begin{figure}[h!]
\centering
\includegraphics[width=1.0\textwidth]{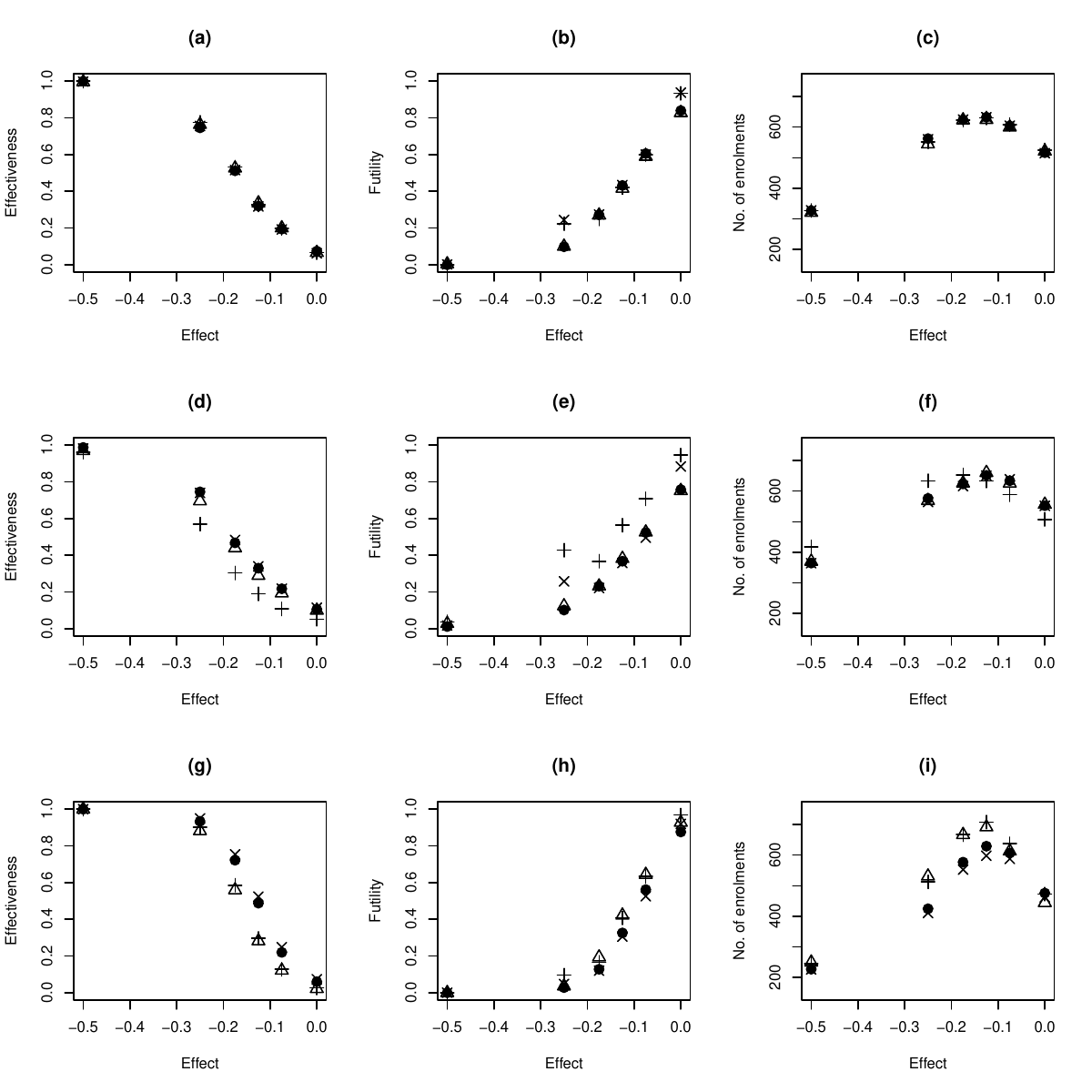}
\caption{\label{fig:trial_suc_fut_n_all}Proportion of trials stopped due to effectiveness (first column) or futility (second column) with the associated average number of enrolments (third column) for when the data were generated based on the exponential (first row), Weibull (second row) and super (third row) models and the treatment effect was estimated based on the exponential ($+$), Weibull ($\times$), general Bayesian ($\bullet$) and piecewise exponential hazard ($\triangle$) models. All results are based on 500 trial simulations.}
\end{figure}

For each simulated trial, the expectation of the effectiveness decision rule was recorded. 
From this, we evaluated the proportion of simulated trials that stopped due to sufficient evidence of effectiveness i.e.\ this expectation being greater than $\Delta_e=0.9$.
These proportions are shown in the first column of Figure \ref{fig:trial_suc_fut_n_all} when the (a) exponential, (d) Weibull and (g) super model generated the data where the different plot characters denote the different analysis models.
From the results, when the exponential model generated the data, the proportion of trials that stopped due to effectiveness was similar across all four analysis models.
When the Weibull model generated the data, the general Bayesian and piecewise exponential models yielded similar proportions to the Weibull model.
The same was not true for the exponential model which typically yielded lower proportions.
When the super model generated the data, the Weibull and general Bayesian models yielded similar proportions, with the proportions for the Weibull model perhaps being slightly larger overall.
This may be linked to the overestimation of the treatment effect observed above.
For the exponential and piecewise exponential models, these generally yielded the lowest proportions, and again this may be linked to these models generally underestimating the effect of treatment.
These results suggest the trial would be less likely to stop for effectiveness (when the treatment is actually effective) if the exponential and piecewise exponential models were adopted for inference (as opposed to the Weibull and general Bayesian models).
The exceptions to this were when the effect of treatment was large and when the treatment did not have an effect (as all proportions were similar in both of these cases).

In each simulated trial, the expectation of the futility decision rule was also recorded.
Similar to the above, we used these expectations to determine the proportion of trials that stopped due to futility i.e.\ this expectation being less than $\Delta_f=0.05$.
These proportions are shown in the second column of Figure \ref{fig:trial_suc_fut_n_all} when the (b) exponential, (e) Weibull and (h) super models generated the data.
As shown, when the exponential model generated the data, these proportions were similar across the four analysis models.
When no treatment effect was assumed, fewer trials stopped for futility under the general Bayesian and piecewise exponential models.
When the Weibull model generated the data and the treatment had an effect, more trials were stopped due to futility under the exponential model, with these proportions being similar between the Weibull, general Bayesian and piecewise exponential models.
However, when there was no effect of treatment, again fewer trials were stopped for futility under the general Bayesian and piecewise exponential models.
When the super model generated the data and the treatment had an effect, the proportion of trials stopped for futility was the lowest under the Weibull and general Bayesian models, and highest for the exponential and piecewise exponential models.
When there was no effect of treatment, the general Bayesian model had the lowest proportion of trials stopped for futility.
Thus, overall it appears as though having an undefined baseline hazard function or forming a non-parametric estimate of this function can lead to fewer trials being stopped for futility when the treatment has no effect.
Otherwise, the Weibull and general Bayesian models generally stopped less often for futility when the treatment had an effect.



Lastly, we assessed the total number of enrolments in each simulated trial.
Figure \ref{fig:trial_suc_fut_n_all} shows the average number of enrolments when the (c) exponential, (f) Weibull and (i) super models generated data.
From the results, when the exponential model generated the data, all four models yielded a similar number of enrolments on average.
When the Weibull model generated the data, all models except the exponential model yielded a similar number of enrolments on average.
For the exponential model, a reduced number of enrolments were observed for small to moderately sized treatment effects, and an increased number of enrolments was observed in the other cases.
When the super model generated the data, a lower number of enrolments were observed on average for the Weibull and general Bayesian model when the treatment had an effect compared to the exponential and piecewise exponential models.
When the treatment had no effect, the general Bayesian model had the largest average number of enrolments which may correspond to this approach being less likely to stop due to futility (as discussed above).
We note that this difference was relatively minor.

\subsection{Adaptive trial simulation under a range of hazard functions}

Lastly, we extended the above simulation study to explore the performance of the four analysis models under a range of hazards functions that could be observed under the super model.  
To specify this range of hazard functions, a uniform prior on $\beta$ was assumed with bounds $-0.75$ and $-0.25$ and the prior on the baseline hazard function was as specified for the super model in the illustrative example.

\begin{figure}[h!]
\centering
\includegraphics[width=0.45\textwidth]{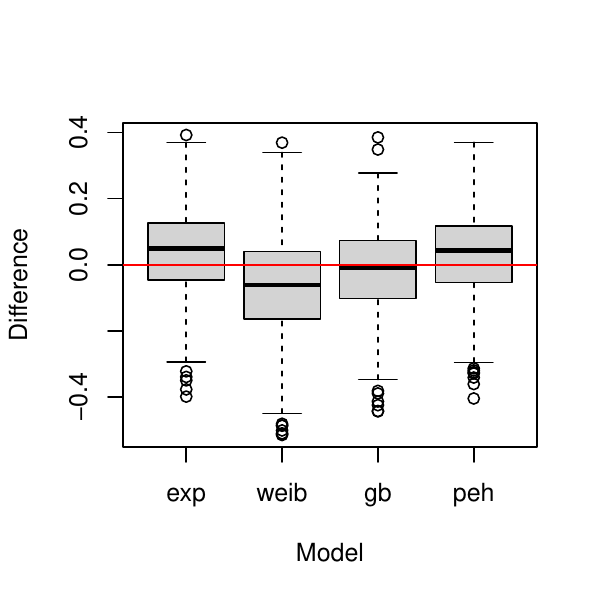}
\caption{\label{fig:range_results}Simulation results from considering a range of hazards functions under the super model where the boxplots show the distribution of the difference between the posterior mode and the assumed treatment effect when the exponential (`exp'), Weibull (`weib'), general Bayesian (`gb') and piecewise exponential (`peh') models were used to estimate the treatment effect. All results are based on 500 trial simulations.}
\end{figure}

A summary of the estimated treatment effects from this simulation study is provided in Figure \ref{fig:range_results} which shows the distribution of the difference between the underlying treatment effects and the posterior modes based on the exponential, Weibull, general Bayesian and piecewise exponential models.  
As can be seen, the exponential, Weibull and piecewise exponential models provided relatively biased estimates of the treatment effect with a median difference of $0.0503$, $-0.0607$ and $0.04283$, respectively.
This agrees with the results above where the exponential and piecewise exponential models underestimated the treatment effect while the Weibull model tended to overestimate it.
In comparison, the general Bayesian model provided a relatively unbiased estimate of the treatment effect with a median difference of $-0.0087$.  
Moreover, if we consider the median squared error, then the exponential, Weibull, general Bayesian and piecewise exponential models yielded $0.0091$, $0.0100$, $0.0084$ and $0.0087$, respectively, providing further support for the use of a general Bayesian model to design adaptive clinical trials with time-to-event outcomes.

For the number of enrolments, the use of the exponential and piecewise exponential models resulted in an average of 261.7 and 268.6 enrolments, respectively.
These were larger than the averages observed under the Weibull and general Bayesian models i.e. 238.5 and 243.7, respectively.
This aligns with results presented above where the exponential and piecewise exponential models generally resulted in more participants being enrolled compared to the Weibull and general Bayesian models when the treatment had an effect.
Also, as seen above, a similar average number of enrolments were observed under the Weibull and general Bayesian models, with only 5.2 fewer participants being enrolled, on average, under the Weibull model.


In terms of power and the proportion of trials stopped due to effectiveness or futility, only minor differences were observed between the four analysis models, so these results have been omitted.

\section{Discussion}\label{sec:conc}

We have proposed an approach to design adaptive clinical trials for time-to-event outcomes within a general Bayesian framework.  
Our motivation for this was to exploit the robustness properties of this inference framework to allow the baseline hazard function to only be implicitly defined {\it a priori}.
Throughout this paper, we have shown that our proposed approach was particularly useful as misspecifying or forming a non-parametric estimate of the baseline hazard function could lead to biased estimates of the treatment effect, lower power and an inappropriate Type I error.
There were also implications for the trial conduct including having a lower chance of stopping due to effectiveness and a higher chance of declaring futility when the treatment had an effect.
In addition, we observed that the number of enrolments under the general Bayesian model was either lower or similar to alternative methods.
This suggests that estimating the treatment effect based on an undefined baseline hazard function did not result in any noticeable loss in estimation efficiency compared to adopting the underlying parametric form or forming a non-parametric estimate.
The only potential drawback that was observed was a reduced chance of declaring futility when the treatment had no effect, and this also appeared to be a limitation of the semiparametric approach considered throughout.
Accordingly, we hope that our proposed approach can be considered and further developed into the future to design adaptive trials where time-to-event outcomes are of primary interest.

Through our motivating clinical trial, we considered decision rules for expected effectiveness and futility.
It is worth noting that our approach could be considered when a variety of other decision rules are being employed.
For example, when multiple treatments are being randomised, the effectiveness rule could be extended to determine the best treatment overall (rather than just in comparison to control).
Similar extensions could also be considered for assessing inferiority and (practical) equivalence.\cite{shih_et_al_2004,christensen_2007}
In these cases, the trial could proceed with the associated treatment arm being dropped from randomisation, as appropriate.
In addition, our approach may also be useful in response-adaptive randomisation where the probability of being allocated certain treatments can change throughout the trial such that the more promising treatments are allocated more often.\cite{robertson_et_al_2023}
Guidance for employing these types of decision rules can also be found in the literature. \cite{granholm_et_al_2023}
Approaches for adjusting the allocation probabilities are typically based on the posterior distribution of the treatment effect, which is readily available in our approach.
The properties of such sequential procedures (based on a general posterior distribution) would be interesting to explore into the future.

A challenge in adopting a general Bayesian approach to re-design the ORVAC trial was that the loss function was not linked to the complete likelihood function.
This meant it was unclear how data could be generated to evaluate decision rules.
To address this, we proposed to use a super model which was specified to be flexible such that it could simulate a wide range of data sets.
Here, this related to specifying a wide range of hazard functions.
However, specifying such a model in practice may be difficult.
Indeed, a model that is too flexible could potentially lead to an inefficient design.
We investigated this for the illustrative example where vaguer prior information was assumed for the super model.
Specifically, we assumed the values of the baseline hazard at each knot were uniformly distributed in $(0,1)$ (as opposed to $(0,0.5)$).
While the power and Type I error results remained robust to these vaguer priors (see Figure SM.1), we suggest that, in general, this may not be the case so careful specification of the super model will be needed.
For example, this may involve close consultation with clinicians to ensure a flexible yet realistic super model (e.g.\ based on the prior predictive distribution) is being proposed.

While the benefits of our approach have been demonstrated, there are limitations of this work.  
One such limitation is the need to assume proportional hazards between treatment groups.  
While this is a common assumption, particularly when designing trials, we suggest that in practice such an assumption may not be appropriate. \cite{WU20182946}
Accordingly, we are interested in extending our approach to relax this assumption, potentially by considering a flexible functional form for the proportionality parameter, and undertaking inference within a general Bayesian framework.  
Another limitation, specific to our proposed general Bayesian approach, is that we focus on estimating hazard ratios.
In clinical trials with time-to-event outcomes, other estimands may be of interest (e.g.\ survival probability or restricted mean survival time) such that the baseline hazard cannot remain undefined.
In such cases, standard Bayesian approaches may be preferable.

Throughout this paper, we evaluated the performance of trial designs via metrics that were averaged over a prior distribution.
For example, in the illustrative example, trial designs were compared via power which was evaluated by averaging power over a range of baseline hazard functions that could be observed under the super model. 
While averaging over a prior distribution is commonplace in Bayesian design,\cite{ChalonerVerdinelli} such an approach may mean that reduced power may be observed for some specific baseline hazard functions.
We have shown this for the illustrative example through considering 10 specific baseline hazards function (shown in Figure SM.8(a)), and re-evaluating power for each of these models where general Bayesian inference was used to estimate the treatment effect (Figure SM.8(b)).
As shown, as sample size increases, an increase in power was observed (similar to that shown in Figure \ref{fig:toy_power_type1}) with some models yielding lower than average power.
This re-evaluation was also undertaken for the ORVAC trial for the same 10 baseline hazard functions where a range of outcomes were observed for the Type I error, effectiveness, futility and the average number of enrolments, see Figure SM.9.
The patterns seen here are similar to those observed in Figure \ref{fig:trial_suc_fut_n_all} when the super model generated the data.
Given the variability in the metrics, a more conservative approach to trial design may be to consider a maximin type approach where, instead of averaging a metric, a minimum level of performance is evaluated and optimised over a range of plausible scenarios e.g.\ over a range of baseline hazard functions based on the super model.\cite{dette_haines_imhof_2007,wu_wong_crespi_2017}
From Figure SM.8(b), such an approach may select a sample size of (say) 900 (instead of 800 from Section \ref{sec:illustrative}) as this would provide greater than 80\% power for all 10 models rather than greater than 80\% power on average across all models.
We note that such a conservative approach was straightforward to implement within our proposed approach to trial design.

More generally, there is scope to extend our proposed approach to other data types including to overdispersed settings.  
In such cases, considering a quasi-likelihood \cite{wedderburn_1974} might be appropriate or an appropriate loss function could be developed, and a general Bayesian approach could be adopted for inference.
An additional challenge of using other loss functions could be the need to calibrate $w$ in Equation (\ref{eq:general_bayes}). \cite{syring_ryan_2018}
Given this will need to be evaluated many times within trial simulation, computationally efficient approaches are needed.
This is an area we plan to explore into the future.





~\vspace{-1.25cm}
\bibliographystyle{ama}
\bibliography{biblio}

\newpage

\begin{algorithm}
\caption{Simulate a Bayesian adaptive clinical trial with time-to-event outcomes}
\label{alg:sim}
\begin{algorithmic}[1]
\STATE Specify trial inputs 
\FOR{$k = 1:K$}
  \STATE Enrol $m_k$ participants. Set $n_k$ to be the current total number of enrolments
  \STATE Randomly assign the $m_k$ new participants their treatment allocations 
  \STATE Simulate observations for the $m_k$ new participants based on $h(t|\beta,\bm{\tau})$
  \IF{planned interim analysis}
    \STATE Find $p(\beta|\bm{y}_{- r_k},\bm{y}^*_{r_k},\bm{x}_{1:n_k})$
    \STATE Evaluate decision rules based on $p(\beta|\bm{y}_{- r_k},\bm{y}^*_{r_k},\bm{x}_{1:n_k})$
    \IF{any decision rule met}
      \STATE Stop or adjust trial
    \ENDIF
  \ENDIF
\ENDFOR
\STATE Find $p(\beta|\bm{y}_{1:N},\bm{x}_{1:N})$ and draw conclusions
\end{algorithmic}
\end{algorithm}

\newpage 

\begin{algorithm}
\caption{Approximation of decision rule for assessing expected treatment effectiveness}
\label{alg:1}
\begin{algorithmic}[1]
\STATE Input: $\bm{y}_{-r_k},\bm{y}^*_{r_k},\bm{x}_{1:n_k}$
    \FOR{$b = 1:B$}
    \STATE $\bm{z}_b \sim p(\bm{z}|\bm{y}_{-r_k},\bm{y}^*_{r_k},\bm{x}_{1:n_k})$~Simulate potentially censored outcomes for participants indexed $r_k$
    \STATE Find $p(\beta|\bm{y}_{-r_k},\bm{z}_b,\bm{x}_{1:n_k})$ The updated posterior distribution of treatment effect
    \STATE $\bm{\Lambda}_b = \mathcal{I}(P(\beta < 0|\bm{y}_{-r_k},\bm{z}_b,\bm{x}_{1:n_k}) > \Delta)$  Evaluate if effectiveness would be concluded
    \ENDFOR
    \STATE $\delta_e = \frac{1}{B}\sum_{b=1}^B \bm{\Lambda}_b$
    \IF{$\delta_e > \Delta_e$}
    \STATE Stop the trial
  \ENDIF
\end{algorithmic}
\end{algorithm}

\newpage


\begin{table}[h!]
    \centering
    \begin{tabular}{l|c|cccccc}
    \hline
     Model & Treatment effect & exp & weib & gb & peh \\
      \hline
      exp   & ~0.000 & 0.0046 & 0.0046 & 0.0046 & 0.0046 \\
            & -0.075 & 0.0047 & 0.0048 & 0.0048 & 0.0047 \\
            & -0.125 & 0.0048 & 0.0048 & 0.0048 & 0.0048 \\
            & -0.175 & 0.0048 & 0.0048 & 0.0048 & 0.0049 \\
            & -0.250 & 0.0048 & 0.0049 & 0.0049 & 0.0049 \\
            & -0.500 & 0.0051 & 0.0052 & 0.0052 & 0.0052 \\
    \hline
      weib  & ~0.000 & 0.0012 & 0.0039 & 0.0039 & 0.0030 \\
            & -0.075 & 0.0021 & 0.0039 & 0.0039 & 0.0033 \\
            & -0.125 & 0.0038 & 0.0039 & 0.0039 & 0.0039 \\
            & -0.175 & 0.0062 & 0.0039 & 0.0038 & 0.0048 \\
            & -0.250 & 0.0111 & 0.0039 & 0.0039 & 0.0060 \\
            & -0.500 & 0.0351 & 0.0039 & 0.0039 & 0.0162    \\
    \hline
      super & ~0.000 & 0.0029& 0.0052& 0.0042& 0.0030 \\
            & -0.075 & 0.0031& 0.0053& 0.0043& 0.0033 \\
            & -0.125 & 0.0034& 0.0054& 0.0043& 0.0035 \\
            & -0.175 & 0.0038& 0.0055& 0.0043& 0.0038 \\
            & -0.250 & 0.0045& 0.0058& 0.0043& 0.0045 \\
            & -0.500 & 0.0077& 0.0076& 0.0045& 0.0076 \\
      \hline
    \end{tabular}
    \caption{Mean squared error of posterior modes for the treatment effect from trial simulation (without interim analyses) when the exponential (`exp'), Weibull (`weib') and super models generated the data (rows) and the treatment effect was estimated based on the exponential, Weibull, general Bayesian (`gb') and piecewise exponential hazard (`peh') models (columns). All results are based on 500 trial simulations.}
    \label{tab:mse_trial}
\end{table}

\newpage

\begin{table}[h!]
    \centering
    \begin{tabular}{l|c|ccccccc}
    \hline
      Model & Treatment effect & exp & weib & gb & peh \\
      \hline
      exp   & 0      & 0.024 &0.022 &0.024 &0.022 \\
            & -0.075 & 0.220 &0.210 &0.214 &0.214 \\
            & -0.125 & 0.456 &0.452 &0.454 &0.472 \\
            & -0.175 & 0.720 &0.718 &0.718 &0.728 \\
            & -0.250 & 0.948 &0.944 &0.944 &0.942 \\
            & -0.500 & 1.000 &1.000 &1.000 &1.000    \\
    \hline
      weib  & 0      & 0.000 &0.032 &0.032 &0.032 \\
            & -0.075 & 0.014 &0.242 &0.242 &0.222 \\
            & -0.125 & 0.086 &0.548 &0.548 &0.514 \\
            & -0.175 & 0.308 &0.800 &0.804 &0.672 \\
            & -0.250 & 0.738 &0.986 &0.986 &0.910 \\
            & -0.500 & 1.000 &1.000 &1.000 &1.000    \\
    \hline
      super & 0      & 0.008 &0.040 &0.030 &0.014 \\
            & -0.075 & 0.164 &0.318 &0.262 &0.158 \\
            & -0.125 & 0.400 &0.572 &0.520 &0.404 \\
            & -0.175 & 0.710 &0.852 &0.812 &0.710 \\
            & -0.250 & 0.944 &0.982 &0.978 &0.942 \\
            & -0.500 & 1.000 &1.000 &1.000 &1.000\\
      \hline
    \end{tabular}
    \caption{Proportion of trial simulations where treatment effectiveness was declared when the exponential (`exp'), Weibull (`weib') and super models generated the data (rows) and the exponential, Weibull, general Bayesian (`gb') and piecewise exponential (`peh') models were used to estimate the treatment effect (columns). All results are based on 500 trial simulations.}
    \label{tab:type11}
\end{table}

\newpage


\begin{table}[h!]
    \centering
    \begin{tabular}{l|c|cccccc}
    \hline
    Model & Treatment effect & exp & weib & gb & peh \\
      \hline
     exp & ~0.000 & 0.0181 &0.0182 &0.0183 &0.0188 \\
         & -0.075 & 0.0199 &0.0205 &0.0204 &0.0209\\
         & -0.125 & 0.0214 &0.0218 &0.0212 &0.0216\\
         & -0.175 & 0.0222 &0.0220 &0.0216 &0.0229\\
         & -0.250 & 0.0208 &0.0209 &0.0206 &0.0214\\
         & -0.500 & 0.0247 &0.0243 &0.0236 &0.0252\\
\hline
    weib & ~0.000 & 0.0067 &0.0134 &0.0132 &0.0097\\
         & -0.075 & 0.0086 &0.0137 &0.0141 &0.0111\\
         & -0.125 & 0.0094 &0.0149 &0.0155 &0.0125\\
         & -0.175 & 0.0101 &0.0154 &0.0155 &0.0133\\
         & -0.250 & 0.0114 &0.0148 &0.0150 &0.0139\\
         & -0.500 & 0.0187 &0.0199 &0.0199 &0.0191\\
\hline
   super & ~0.000 & 0.0091& 0.0171& 0.0149& 0.0097 \\
         & -0.075 & 0.0107& 0.0193& 0.0161& 0.0116 \\
         & -0.125 & 0.0105& 0.0202& 0.0164& 0.0110 \\
         & -0.175 & 0.0102& 0.0195& 0.0160& 0.0109 \\
         & -0.250 & 0.0094& 0.0193& 0.0150& 0.0098 \\
         & -0.500 & 0.0144& 0.0229& 0.0176& 0.0147 \\
\hline
    \end{tabular}
    \caption{Mean squared error of posterior modes for the treatment effect from adaptive trial simulation (with interim analyses) when the exponential (`exp'), Weibull (`weib') and super models generated the data (rows) and the treatment effect was estimated based on the exponential, Weibull, general Bayesian (`gb') and piecewise exponential hazard (`peh') models (columns). All results are based on 500 trial simulations.}
    \label{tab:mse_adaptive_trial}
\end{table}

\newpage

\begin{table}[h!]
    \centering
    \begin{tabular}{l|c|ccccccc}
    \hline
      Model & Treatment effect & exp & weib & gb & peh \\
      \hline
      exp   & 0      & 0.022& 0.018& 0.030& 0.032 \\
            & -0.075 & 0.174& 0.176& 0.156& 0.168 \\
            & -0.125 & 0.318& 0.340& 0.336& 0.336 \\
            & -0.175 & 0.524& 0.518& 0.518& 0.524 \\
            & -0.250 & 0.770& 0.770& 0.778& 0.770 \\
            & -0.500 & 0.978& 0.974& 0.980& 0.968    \\
    \hline
      weib  & 0      & 0.004& 0.034& 0.036& 0.034 \\
            & -0.075 & 0.046& 0.174& 0.150& 0.183 \\
            & -0.125 & 0.144& 0.338& 0.352& 0.339 \\
            & -0.175 & 0.324& 0.512& 0.500& 0.487 \\
            & -0.250 & 0.578& 0.750& 0.748& 0.699 \\
            & -0.500 & 0.938& 0.974& 0.972& 0.948    \\
    \hline
      super & 0      & 0.024& 0.066& 0.052& 0.020 \\
            & -0.075 & 0.160& 0.296& 0.252& 0.158 \\
            & -0.125 & 0.388& 0.534& 0.522& 0.400 \\
            & -0.175 & 0.656& 0.774& 0.754& 0.647 \\
            & -0.250 & 0.916& 0.926& 0.908& 0.916 \\
            & -0.500 & 0.996& 0.998& 0.998& 0.996\\
      \hline
    \end{tabular}
    \caption{Proportion of adaptive trial simulations where treatment effectiveness was declared when the exponential (`exp'), Weibull (`weib') and super models generated the data (rows) and the exponential, Weibull, general Bayesian (`gb') and piecewise exponential (`peh') models were used to estimate the treatment effect (columns). All results are based on 500 trial simulations.}
    \label{tab:type12}
\end{table}

\end{document}